\documentstyle[aps]{revtex}
\def\ltap{\raisebox{-.55ex}{\rlap{$\sim$}} \raisebox{.4ex}{$<$}}
\def\gtap{\raisebox{-.55ex}{\rlap{$\sim$}} \raisebox{.4ex}{$>$}}
\def\gsim{\mathrel{\gtap}}
\def\lsim{\mathrel{\ltap}}
\def\e{\mbox{e}}

\begin{document}
\flushbottom
\twocolumn[\hsize\textwidth\columnwidth\hsize\csname
@twocolumnfalse\endcsname

\title{Statistics of clustering of ultra-high energy cosmic rays
and the number of their sources}

\author{S.~L.~Dubovsky$^a$, P.~G.~Tinyakov$^{a,b}$ and I.~I.~Tkachev$^c$}
\address{
     $^a$Institute for Nuclear Research, 60th October Anniversary
prospect 7a, 117312 Moscow, Russia  \\
     $^b$Institute of Theoretical Physics,
University of Lausanne, CH-1015 Lausanne, Switzerland\\
     $^c$Institute for Theoretical Physics, ETH-H\"onggerberg, CH-8093, Zurich, Switzerland }

\maketitle

\begin{abstract}
Observation of clustering of ultra-high energy cosmic rays (UHECR)
suggests that they are emitted by compact sources. Assuming small
($<3^{\circ}$) deflection of UHECR during the propagation, the
statistical analysis of clustering allows to estimate the spatial
density of the sources $h_*$, including those which have not yet been
observed directly. When applied to astrophysical models involving
extra-galactic sources, the estimate based on 14 events with energy
$E>10^{20}$~eV gives $h_*\sim 6\times 10^{-3}$~Mps$^{-3}$.  With
increasing statistics, this estimate may lead to exclusion of the
models which associate the production of UHECR with exceptional
galaxies such as AGN, powerful radio-galaxies, dead quasars, and 
models based on gamma ray bursts.
\end{abstract}

\vskip2pc]

\section{Introduction} 
Recent analysis of arrival directions of ultra-high energy cosmic rays
(UHECR) reveals groups of events (clusters) with arrival directions
lying within $\sim 3^{\circ}$, the typical angular resolution of the
experiment. The set of 92 observed events with energy $E>4\times
10^{19}$~eV contains 7 doublets and 2 triplets \cite{uchihori}. The
small probability of chance coincidence, of the order of $10^{-3}$
\cite{AGASA,uchihori}, suggests that clustering is a result of the
existence of compact sources. At higher energies, $E > 10^{20}$~eV, one
doublet out of 14 events is observed.

Compact sources of UHECR are naturally explained in astrophysical
models where they are associated with possible UHECR production sites,
such as AGN~\cite{agn}, hot spots of powerful
radio-galaxies~\cite{hotspots}, dead quasars~\cite{deadquasars} and
gamma-ray bursts (GRB)~\cite{grb}. These models have much in
common. They assume that primary particles are protons; the sources of
the observed UHECR have, therefore, to lie within the GZK cutoff
\cite{GZK} distance. For energies $E \approx 10^{20}$~eV the GZK radius is
$R_{\rm GZK} \sim 50 $~Mpc, while at $E \gsim 2\times 10^{20}$~eV it
drops to $\sim 20 $~Mpc \cite{Lee}. In all these models the
distribution of sources in space within the GZK sphere is close to
uniform, while the distribution in luminosity does not depend on space
and peaks around a certain value.

An important common feature of these models is a small local density
of sources. The number density of dead quasars is estimated as $\sim
10^{-4}$~Mpc$^{-3}$ \cite{deadquasars}; the number of AGN is $\sim
10\%$ of the number of galaxies \cite{galaxies}, which gives $\sim
5\times 10^{-4}$~Mpc$^{-3}$. Most likely, only a small fraction of
them is capable of producing UHECR with energies $E > 10^{20}$~eV. In
the case of GRB the effective density of sources is determined by the
rate $\gamma$ of GRB and the typical time delay $\tau$ of UHECR
particles. Taking $\tau \lsim 10^5$~yr and the rate $\gamma \sim
2\times 10^{-10}h^3$~Mpc$^{-3}$~yr$^{-1}$ \cite{grbrate} gives the
density of sources $\sim 10^{-5}$~Mpc$^{-3}$.

The purpose of this letter is to show that the observed clustering
favors larger density of sources, provided the propagation of UHECR
with energy $E > 10^{20}$~eV is not strongly affected by
extra-galactic magnetic fields. The latter assumption is justified if
the existing bound on extra-galactic magnetic field $B\lsim 10^{-9}$~G
\cite{Bbound} is valid.

\section{Statistics of clustering} 

The observable quantities which characterize clustering are $\bar
N_m$, the expected numbers of clusters of different multiplicities
$m$ (e.g., $\bar N_1$ and $\bar N_{2}$ are the expected
numbers of single and double events, respectively).
They depend on the total exposure of the experiment $B$ and the
distribution of sources in the flux they produce%
\footnote{Here and below we mean the integral flux of cosmic rays with
energies above some energy threshold. It measures the average number of events
per unit time per unit area of the detector.},
$n(F)$, which is defined in such a way that the number of sources with
the flux from $F$ to $F+dF$ is $dS = n(F)dF$.  The events which
come from the same source at different times are statistically
independent and therefore have the Poisson distribution. Thus, the
expected number of clusters is
\begin{equation}
\bar N_m = \int_0^{\infty} {(F B)^m\over m!} \e^{-F B} n(F) dF. 
\label{N_m}
\end{equation}
This equation implies that the expected total number of events $N_{\rm
tot}$ is
\begin{equation}
\bar N_{\rm tot}= \sum_m m \bar N_m 
= B\int_0^{\infty} F n(F)dF = BF_{\rm tot},
\label{Ntot}
\end{equation} 
as it should be. The probability to observe $k$ clusters of
multiplicity $m$ is also given by the Poisson distribution,
\begin{equation}
P_m(k) = {(\bar N_m)^k\over k!}\e^{-\bar N_m}.
\label{poisson}
\end{equation}

Any model of UHECR can be characterized by the distribution of sources
in distance and luminosity $f(r,L)$ (in the case of anisotropic distribution
it should be understood as average over the sphere, $f(r,L) \equiv
\int f({\bf r},L) d\Omega/4\pi)$). In order to express $n(F)$ and $\bar N_m$
in terms of the distribution function $f(r,L)$, consider the sources
at distances from $r$ to $r+dr$. The number of such sources with
luminosities from $L$ to $L+dL$ is
\begin{equation}
dS = f(r,L)\, 4\pi r^2 dr\, dL. 
\label{dS}
\end{equation}
Making use of the relation $F = {L/ 4\pi r^2}$ and integrating over
$r$ one finds $dS=n(F)dF$, where 
\begin{equation}
n(F) = (4\pi)^2 \int_0^{\infty} dr\,r^4 f(r,4\pi r^2F).   
\label{N[L]}
\end{equation}
Here we have neglected the curvature effects since they are small at
distances of order 50~Mpc.

In the case of the astrophysical models, the distribution function
$f(r,L)$ is uniform in space and depends only on the luminosity. The
GZK effect, however, makes distant sources fainter by a factor
$\exp(-r/R)$. This is equivalent to substituting 
\begin{equation}
f(r,L) \rightarrow \e^{r/R} h(L\e^{r/R})
\label{fGZK}
\end{equation}
in Eq.(\ref{N[L]}), where $h(L)$ is the distribution of sources in the
luminosity. The exact value of $R$ can be determined by numerical
simulations of UHECR propagation with full account of the energy
dependence. For protons with $E>10^{20}$~eV the simulations give
$R\simeq 25$~Mpc
\cite{DKTT}.

\section{Number of sources}

A key parameter which enters the distribution $f(r,L)$ is the
normalization, or the spatial density of sources, which can be
characterized by the number of sources $S$ within the sphere of a
radius $R$. An important information about $S$ can be obtained from
statistical analysis of clustering even if the functional form of the
distribution $f(r,L)$ is not known. The idea is to find the
distribution $f(r,L)$ which corresponds to the {\em minimum} number of
sources $S_*$ with total number of events, $\bar{N}_{\rm tot}$, and
the number of events in clusters, $\bar N_{\rm cl} \equiv \bar N_{\rm
tot} - \bar N_1$, being fixed. We will show in a moment that in the
case $\bar N_{\rm cl}\ll\bar N_{\rm tot}$ the number $S_*$ is
surprisingly large, much larger than the number of the sources already
observed.

It is intuitively clear why in the case $\bar N_{\rm cl}\ll \bar
N_{\rm tot}$ the number of sources is much larger than $\bar N_{\rm
tot}$ \cite{PW}. In order to produce $\sim N_{\rm tot}$ single events
by $\sim N_{\rm tot}$ sources each of them has to be bright
enough. But then a large number of doublets would be produced as well.
Since this is not the case, i.e. most of the resolved sources are dim
and produce at most one event, one concludes that there is a large
number of sources which have not yet revealed themselves.  Assuming
that all sources have the same flux $F$ one finds from Eq. (\ref{N_m})
$\bar N_1\sim S\bar n$ and $\bar N_2\sim S\bar n^2/2$, where $\bar
n=FB$ is the average number of events produced by one source.
Therefore, $S\sim \bar N_1^2/(2\bar N_2)\sim\bar N_{\rm tot}^2 /\bar
N_{\rm cl}$, i.e. much larger than $\bar N_{\rm tot}$. Using methods
described in the Appendix it is possible to show that the case of
equal fluxes corresponds to the absolute minimum of S. However, this
distribution is unphysical.  Many realistic situations correspond to a
homogeneous distribution of sources in space when more distant sources
are fainter; consequently, their number has to be even larger than
predicted by the above estimate.

In astrophysical models the distribution $f(r,L)$ is given by
Eq. (\ref{fGZK}) containing one unknown function $h(L)$. The minimum
density of sources is determined by minimizing over $h(L)$. As is
shown in the Appendix, the minimum is reached at the delta-function
distribution
\begin{equation}
h(L) = h_* \delta(L-L_*), 
\label{h=delta}
\end{equation}
where $L_*$ is the luminosity of the sources and $h_*$ is their spatial
density.  The unknown parameters $h_*$ and $L_*$ can be related to
$\bar N_{\rm tot}$ and $\bar N_{\rm cl}$ by making use of
Eqs.(\ref{N_m}) and (\ref{Ntot}). Introducing the notations
\begin{eqnarray}
S_*  &=& (4\pi/3)  R^3 h_*, 
\label{S*}\\ \nonumber
\nu_* &=& B L_*/(4\pi R^2),
\end{eqnarray}
where $\nu_*$ is the expected number of
events from one source at the distance $R$ in the absence of the GZK
cutoff, one has the following equations,
\begin{eqnarray} 
\label{eq1}
\bar N_{\rm tot} &=& 3 S_* \nu_* \; ,\\
\label{eq2}
\bar N_1 &=& 3S_*\nu_* \int_0^{\infty} dx \exp 
\left(- x - \nu_*x^{-2}\e^{-x} \right).
\end{eqnarray}
These equations can be solved perturbatively at small
$\bar N_{\rm cl} \ll \bar N_{\rm tot}$. One finds
\begin{equation}
\nu_*  \simeq {1\over\pi}  {\bar N_{\rm cl}^2\over \bar N_{\rm tot}^2 },
\label{nu*}
\end{equation}
\begin{equation}
S_*  \simeq {\pi\over 3} {\bar N_{\rm tot}^3\over\bar  N_{\rm cl}^2 }.
\label{Sfinal}
\end{equation}
If $\bar N_{\rm cl}\ll \bar N_{\rm tot}$, the minimum number of
sources $S_*$ is indeed much larger than $\bar N_{\rm tot}$ and,
therefore, is much larger than the number of sources already
observed. From Eq. (\ref{nu*}), each source produces much less than 1
event in average.

\section{Discussion} 

Let us apply these arguments to the observed events with energies
$E>10^{20}$~eV. In this case, $N_{\rm tot} = 14$ and $N_{\rm cl} = 2$.
The solution to Eqs.(\ref{eq1}) and (\ref{eq2}) is $S_*  \sim 400$, 
which at $R=25$~Mpc corresponds to the density 
\begin{equation}
h_*  \sim 6\times 10^{-3} \quad {\rm Mpc}^{-3}.
\label{Smin-astro}
\end{equation}
This number is large as compared to the density of sources in most of
astrophysical models. However, it should be interpreted with care. One
may expect large statistical fluctuations because both $N_{\rm tot}$
and $N_{\rm cl}$ are small and may not coincide with their expected values.

In order to address this issue quantitatively, let us find the model
which has the largest probability $p(h_*)$ to reproduce the observed
clustering at fixed density of sources $h_*$. To this end, consider
the set of models which are described by Eqs.(\ref{fGZK}) and
(\ref{h=delta}) and are characterized by two parameters $h_*$ and
$L_*$. At fixed density of sources, there remains a freedom of
changing $L_*$. The probability to reproduce the observed data is
maximum for some $L_*$; this probability is $p(h_*)$. By construction,
there are no models with the density of sources smaller than $h_*$, in
which the probability to reproduce the observed data is larger than
$p(h_*)$.

\medskip
\begin{center}
\begin{tabular}{||c||c|c||c|c||c|c||}\hline
proba- & \multicolumn{2}{c||}{14 events}  & \multicolumn{2}{c||}{30
events} & \multicolumn{2}{c||}{60 events}\\ 
bility      & \multicolumn{2}{c||}{1 doublet }  
& \multicolumn{2}{c||}{1 doublet} & \multicolumn{2}{c||}{1 doublet}
 \\ 
$p$ & $h_*$ & $\nu$ & $h_*$ & $\nu$ & $h_*$ & $\nu$ \\ \hline
0.1    & $23$ & 0.51 & $320$ &0.065 & $3100$ & 0.012\\
0.01   & $3.2$ & 5.7  & $63$ & 0.38 & $690$  & 0.058\\
0.001  &    &      & $21$ & 1.3  &  $260$  & 0.15\\ \hline
\end{tabular} 
\end{center}\medskip \nopagebreak
{\small\noindent TABLE~I.~Minimum density of sources $h_*$ 
in the units of $10^{-5}$ Mpc$^{-3}$, and
corresponding source luminosity in the units of $\nu=BL_*/(4\pi R^2)$,
which are required to reproduce the observed clustering with given
probability $p$ for the real experimental data (1 doublet out of 14
events) and for two hypothetical data sets with larger number of
events (one doublet out of 30 events and one doublet out of 60
events).}
\medskip

There is some ambiguity in defining what is ``to reproduce the
observed data''. In the case at hand we request that the number of
singlets is 12 or larger, the number of doublets is 1 or smaller, and
the number of clusters with the multiplicity 3 and larger is
zero. Eq.~(\ref{poisson}) determines the probability $p(h_*,L_*)$ of
such clustering as a function of two parameters $h_*$ and $L_*$. The
probability $p(h_*)$ is found by maximizing $p(h_*,L_*)$ at fixed
$h_*$.  We have performed this calculation numerically. The results
are summarized in Table~1 in the form of lower bounds on the density
of sources. We also present the source luminosity in the units of
$\nu=BL_*/(4\pi R^2)$, i.e. the number of events from a single source
at the distance $R$.

The models where the observed clustering occurs with probability 1\%
have minimum $\sim 2$ sources inside the 25 Mpc sphere. In the latter
case most of the observed 14 events are produced by the sources which
are further than 25 Mpc and thus have to be bright enough. This is
reflected in Table~1 from which we see that these sources would
produce in average $\sim 6$ events each (in the absence of the GZK
cutoff) if placed at 25 Mpc.

It is worth noting that the numbers of Table~1 correspond to the
extreme situation when the distribution of sources is given by
Eq. (\ref{h=delta}) with a particular value of $L_*$. In realistic
models, the distribution of sources in luminosity is usually spread
over an order of magnitude at least. There may also be constraints on
the luminosity of the sources. In these cases, the lower bounds on the
number of sources are higher than in Table~1.

When the new large-area detectors like the Pierre Auger array
\cite{auger} will start operating, the number of observed events will
increase and the statistical errors in determination of the density of
sources will go down. Correspondingly, the lower bounds on the density
will become higher. To show that the bounds may become very high when the
total number of events is still small, we have performed calculations
for two hypothetical situations, 1 doublet out of 30 events, and 1
doublet out of 60 events. The results are also listed in Table~1. The
bounds grow roughly like cube of the number of events, in agreement
with Eq.(\ref{Sfinal}).

To summarize, the statistical analysis of clustering may provide tight
constraints on astrophysical models of UHECR when the number of
clusters is small. In this situation, a key quantity is the density of
sources which can be bound from below in a model-independent way. The
bound grows very fast with the number of single events above $E=
10^{20}$~eV and is potentially dangerous for astrophysical models
which associate production of UHECR with GRB or exceptional galaxies
such as AGN, powerful radio-galaxies and dead quasars. 

Our method equally applies to models in which UHECR are produced in
the Galactic halo, or in which primary particles are immune to the
background radiation. The relation (\ref{Sfinal}) remains valid with a
different numerical coefficient of order one and different meaning of $S_*$.
In the first case $S_*$ is the number of sources in the halo and
detailed analysis shows that statistical properties of clustering of UHECR
are compatible with clumpiness of super-heavy dark matter in decays of
which UHECR may be produced. In the
second case our method counts the number of UHECR sources within the
cosmological horizon, which is inaccessible by other means.

\section*{Acknowledgments}
{\tolerance=400 The authors are grateful to K.A.~Postnov for the
references concerning the number of galaxies and to V.A.~Ru\-ba\-kov
and M.E.~Shaposhnikov for valuable comments and
discussions. This work is supported in part by the Swiss Science
Foundation, grant 21-58947.99. S.D. acknowledges the hospitality of the
University of Lausanne and  was supported in part by RFBR grant
99-02-18410, by the RAS JRP grant 37, and by ISSEP.

}

\section*{APPENDIX: minimum number of sources}

Consider the problem in general terms. First note that by changing
the integration variable in Eq.~(\ref{N[L]}) one can show that any
distribution is equivalent to a factorizable one. So, let us take the
distribution of sources in the form
\begin{equation}
f(r,L) = g(r)h(L).
\label{a:f=g*h}
\end{equation}
Let us fix $g(r)$ and minimize the number of sources
\[
S = 4\pi \int_0^{\infty} r^2 g(r) dr  
\int_0^{\infty} h(L) dL
\]
with respect to the distribution $h(L)$ under the constraints
fixing $\bar N_{\rm tot}$ and $\bar N_1$,
\begin{eqnarray}
\label{a:constr1}
B \int_0^{\infty}g(r) dr 
\int_0^{\infty} Lh(L)dL &=&  \bar{N}_{\rm tot},\\
B \int_0^{\infty} dr dL \; L h(L) 
g(r) \exp \left(-{BL\over 4\pi r^2}\right) &=& \bar{N}_1.
\label{a:constr2}
\end{eqnarray}
This is equivalent to minimizing the functional 
\[
W = 4\pi \int_0^{\infty} dL h(L) 
\int_0^{\infty}g(r) dr  \Biggl\{ r^2 
+ \lambda {BL\over 4\pi} 
\]
\begin{equation}
- \mu {BL\over 4\pi}
\exp \left(-{BL\over 4\pi r^2}\right)  \Biggr\} 
- \lambda \bar{N}_{\rm tot} + \mu \bar{N}_1
\label{a:W}
\end{equation}
with respect to $h(L)$. Here $\lambda$ and $\mu$ are the Lagrange
multipliers. 

The functional (\ref{a:W}) is linear in $h(L)$; denote the coefficient
by $G(L)$,
\[
G(L) = \int_0^{\infty}g(r) dr  \Biggl\{ r^2 
+ \lambda {BL\over 4\pi} 
- \mu {BL\over 4\pi}
\exp \left(-{BL\over 4\pi r^2}\right)  \Biggr\}. 
\] 
At those values of $L$ where $G(L)$ is negative, the minimum of $W$ is
at $h(L)\to \infty$. The latter, however, is not compatible with
Eqs.(\ref{a:constr1}) and (\ref{a:constr2}). Therefore, at the minimum the
values of $\lambda$ and $\mu$ have to be such that $G(L)$ is
non-negative.

At those values of $L$ where $G(L)$ is positive, the minimum of $W$ is
reached at $h(L)=0$. If $G(L)$ is positive at all $L$, then $h(L)$ is
identically zero and Eqs.(\ref{a:constr1}) and (\ref{a:constr2}) are
again violated. Therefore, $\lambda$ and $\mu$ must be such that
$G(L)$ touches zero at some $L_*$. The function $h(L)$ is non-zero
only at this point. Thus, the minimum number of sources corresponds to
the situation when all of them have the same luminosity $L_*$, and we
arrive at the delta-function distribution, Eq.(\ref{h=delta}).

It remains to show that for a given positive function $g(r)$
satisfying $\int g(r) dr<\infty$ the Lagrange multipliers $\lambda$
and $\mu$ can always be chosen in such a way that $G(L)$ is positive
everywhere except an isolated point. To this end rewrite $G(L)$ in the
following form,
\begin{equation}
G(L) = C + \lambda F(L),
\label{a:tmp1}
\end{equation}
where $C=\int r^2 g(r) dr$ is a positive constant and the function
$F(L)$ depends only on the ratio $\mu/\lambda$,
\[
F(L)={BL\over 4\pi} 
\int_0^{\infty}g(r) dr  \left\{ 1 - {\mu\over \lambda} 
\exp \left(-{BL\over 4\pi r^2}\right) \right\}.  
\]
The behavior of the function $F(L)$ is the following. At $L\to 0$ it
goes to zero. At small $L$ it is negative if $\mu/\lambda > 1$ and
positive otherwise. At $L\to\infty$ it grows linearly with $L$, the
coefficient being $B/4\pi \, \int g(r)dr>0$. Therefore, at
$\mu/\lambda > 1$ the function $F(L)$ must have an absolute minimum at
some $L_*>0$ (which is a function of $\mu/\lambda$). Then it is clear
from Eq.(\ref{a:tmp1}) that by choosing $\lambda=-C/F(L_*)> 0$ one can
set $G(L)$ to zero in that particular point. The argument can be
easily generalized to the case of infinite number of sources, $\int
g(r)r^2 dr = \infty$.

In order to apply this argument to the case of astrophysical models,
one should find the factorizable distribution $\tilde f(r,L)$ which
produces the same $n(F)$ as Eq.(\ref{fGZK}). This can be done by
substituting Eq.(\ref{fGZK}) into Eq.(\ref{N[L]}) and changing the
integration variable according to
\begin{equation}
r^2 \exp(r/R) = x^2.
\label{a:var-ch}
\end{equation}
The result reads
\[
\tilde f(x,L) = g(x)h(L),
\]
where 
\[
g(x)=(1+r(x)/2R)^{-1} \e^{-3r(x)/2R}
\]
and $r(x)$ is defined by Eq.(\ref{a:var-ch}).

\end{document}